\documentclass[aps, prd, onecolumn, tightenlines, notitlepage, superscriptaddress, nofootinbib, preprintnumbers, floatfix,showkeys,11pt,altaffilletter]{revtex4-2}

\usepackage{upgreek}

\usepackage[official]{eurosym}
\usepackage[normalem]{ulem}
\usepackage{amstext}
\usepackage{graphicx}
\graphicspath{{figures}}
\usepackage{url}
\usepackage{color}
\usepackage{ulem}
\usepackage[version=4]{mhchem}
\usepackage[utf8]{inputenc}
\usepackage{fontawesome}
\usepackage{yfonts}
\usepackage{slashed}

\pdfoutput=1
\usepackage{textcomp}
\usepackage{comment}
\usepackage{yfonts}
\usepackage{epsfig,amsfonts,mathrsfs,graphicx,color,slashed,multirow}
\usepackage{latexsym,graphicx,slashed,color,enumerate,url,cancel,gensymb}
\usepackage{textcomp}

\usepackage[x11names]{xcolor}
\usepackage[colorlinks,pdfstartview=FitV,breaklinks=true]{hyperref}
\usepackage{booktabs}
\usepackage{adjustbox}
\usepackage{textgreek} 
\usepackage{caption, subcaption} 
\usepackage{booktabs} 
\usepackage{float}

\usepackage{lmodern}
\usepackage{ae,aecompl}
\usepackage{appendix}
\usepackage{orcidlink}
\usepackage{nccmath} 
\usepackage{textcomp}
\allowdisplaybreaks 


\def\cevns{CE\textnu NS}
\def\eves{E\textnu ES}
\def\d{\mathrm{d}}
\newcommand{\qtransfer}{\left|\mathbf{q}\right|}

\definecolor{vdrgreen}{rgb}{0.0, 0.6, 0.0}

\definecolor{byzantium}{rgb}{0.44, 0.16, 0.39}

\makeatletter
    \newcommand{\colorboxed}[3][white]{\fcolorbox{#2}{#1}{\m@th$\displaystyle#3$}}
\makeatother

\AtBeginDocument{\hypersetup{citecolor=byzantium,linkcolor=byzantium,urlcolor=byzantium}}
\usepackage{appendix}

\begin{document}

\title{{\LARGE Neutrino electromagnetic properties and sterile dipole portal in light of the first solar \cevns~data}}

\author{V. De Romeri~\orcidlink{0000-0003-3585-7437}}
\email{deromeri@ific.uv.es}
\affiliation{Instituto de F\'{i}sica Corpuscular (CSIC-Universitat de Val\`{e}ncia), Parc Cient\'ific UV C/ Catedr\'atico Jos\'e Beltr\'an, 2 E-46980 Paterna (Valencia), Spain}

\author{D. K. Papoulias~\orcidlink{0000-0003-0453-8492}}\email{dipapou@ific.uv.es}
\affiliation{Instituto de F\'{i}sica Corpuscular (CSIC-Universitat de Val\`{e}ncia), Parc Cient\'ific UV C/ Catedr\'atico Jos\'e Beltr\'an, 2 E-46980 Paterna (Valencia), Spain}

\author{G. Sanchez Garcia~\orcidlink{}}\email{gsanchez@ific.uv.es}
\affiliation{Instituto de F\'{i}sica Corpuscular (CSIC-Universitat de Val\`{e}ncia), Parc Cient\'ific UV C/ Catedr\'atico Jos\'e Beltr\'an, 2 E-46980 Paterna (Valencia), Spain}
\affiliation{Departament de F\'isica  Te\'orica,  Universitat  de  Val\`{e}ncia, Spain}

\author{C. A. Ternes~\orcidlink{0000-0002-7190-1581}}
\email{christoph.ternes@lngs.infn.it}
\affiliation{Istituto Nazionale di Fisica Nucleare (INFN), Laboratori Nazionali del Gran Sasso, 67100 Assergi, L’Aquila (AQ), Italy
}
\author{M. T\'ortola~\orcidlink{}}\email{mariam@ific.uv.es}
\affiliation{Instituto de F\'{i}sica Corpuscular (CSIC-Universitat de Val\`{e}ncia), Parc Cient\'ific UV C/ Catedr\'atico Jos\'e Beltr\'an, 2 E-46980 Paterna (Valencia), Spain}
\affiliation{Departament de F\'isica  Te\'orica,  Universitat  de  Val\`{e}ncia, Spain}


\begin{abstract}
Despite being neutral particles, neutrinos can acquire non-zero electromagnetic properties from radiative corrections that can be induced by the presence of new physics. Electromagnetic neutrino processes induce spectral distortions in neutrino scattering data, which are especially visible at experiments characterized by low recoil thresholds. We investigate how neutrino electromagnetic properties confront the recent indication of coherent elastic neutrino-nucleus scattering (\cevns) from $^8$B solar neutrinos in dark matter direct detection experiments. We focus on three possibilities: neutrino magnetic moments, neutrino electric charges, and the active-sterile transition magnetic moment portal. We analyze recent XENONnT and PandaX-4T data and infer the first \cevns-based constraints on electromagnetic properties using solar $^8$B neutrinos.
\end{abstract}
\maketitle


\section{Introduction}
The observation of neutrino oscillations~\cite{Kajita:2016cak,McDonald:2016ixn} suggests that the Standard Model (SM) of particle physics must be extended to account for small neutrino masses. Even though neutrinos are electrically neutral with no tree-level interactions with the electromagnetic field, they may obtain non-zero couplings to photons via quantum loop effects in some of these SM extensions. Within the SM, the only electromagnetic properties that neutrinos can acquire are tiny charge radii, also arising due to radiative corrections~\cite{Bernabeu:2000hf,Bernabeu:2002nw,Bernabeu:2002pd}. For a detailed review of neutrino electromagnetic properties, we refer the reader to Refs.~\cite{Giunti:2014ixa,Giunti:2024gec}.

One possibility of probing neutrino electromagnetic properties is through their phenomenological implications on neutrino scattering off electrons or nuclei. In addition to traditional neutrino experiments, facilities dedicated to the direct search of dark matter (DM) particles have been proven to be sensitive to large fluxes of astrophysical neutrinos through elastic neutrino-electron scattering (\eves)~\cite{Monroe:2007xp,Vergados:2008jp,Strigari:2009bq,Billard:2013qya} or coherent elastic neutrino-nucleus scattering (\cevns)~\cite{Freedman:1973yd,Abdullah:2022zue}. 
With current detector technologies, only the $pp$ and $^7$Be components of the solar neutrino flux produce detectable \eves~rates, while the detection of \cevns~is due to $^8$B solar neutrinos, which require a very low energy threshold.
Sensitivity to such low-energy recoils has recently been achieved by low-threshold dual-phase liquid xenon detectors (LXe), namely,  the experiments XENONnT~\cite{XENON:2020gfr,XENON:2024ijk} and PandaX-4T~\cite{PandaX:2022xqx,PandaX:2024muv}, giving rise to a \textit{neutrino fog} that presents important challenges for DM searches~\cite{OHare:2021utq}. On the other hand, both \eves~and \cevns~signals at DM facilities offer a new possibility to test the neutrino sector, including the non-trivial electromagnetic properties of neutrinos; see, for example,~\cite{Harnik:2012ni,Cerdeno:2016sfi,Dutta:2017nht,Gelmini:2018gqa,Essig:2018tss,AristizabalSierra:2020edu,Boehm:2020ltd,AristizabalSierra:2020edu,AristizabalSierra:2020zod,Amaral:2020tga,Dutta:2020che,Suliga:2020jfa,deGouvea:2021ymm,Amaral:2021rzw,AtzoriCorona:2022jeb,A:2022acy,XENON:2022ltv,Khan:2022bel,Coloma:2022umy,Aalbers:2022dzr,Alonso-Gonzalez:2023tgm,Giunti:2023yha,Amaral:2023tbs,Majumdar:2024dms,Demirci:2023tui,DeRomeri:2024dbv,Demirci:2024vzk}. 
Indeed, these new results~\cite{XENON:2024ijk,PandaX:2024muv} have already motivated phenomenological studies for new physics searches including non-standard neutrino interactions (NSI)~\cite{AristizabalSierra:2024nwf,Li:2024iij}, light mediators~\cite{Xia:2024ytb,DeRomeri:2024iaw,Blanco-Mas:2024ale}, and also the determination of the weak mixing angle at low momentum transfer~\cite{Maity:2024aji,DeRomeri:2024iaw}.

In this paper, we investigate potential neutrino electromagnetic properties by analyzing recent data collected by XENONnT~\cite{XENON:2024ijk} and PandaX-4T~\cite{PandaX:2024muv}, from \cevns~induced by $^8$B solar neutrinos scattering off xenon nuclei in their detectors.
If neutrinos interact with photons through radiative corrections, then the \cevns~cross section would be altered, giving rise to spectral distortions at low recoil energies. Thus, analyzing the spectra observed in DM experiments allows us to place complementary constraints on neutrino electromagnetic interactions.
Our analysis focuses on neutrino magnetic moments, neutrino millicharges, and an active-to-sterile magnetic moment portal. For these scenarios, we obtain the first-ever constraints from $^8$B solar \cevns~and  compare them with existing bounds.
In this context, we also present the first evaluation of the \eves~constraint on the sterile dipole portal through a combined analysis of data collected by the experiments XENONnT~\cite{XENON:2022ltv}, LUX-ZEPLIN (LZ)~\cite{LZ:2022lsv}, and PandaX-4T~\cite{PandaX:2022ood}. 

The paper is organized as follows. In Sec.~\ref{sec:theory}, we discuss the cross sections relevant to the analysis of non-standard neutrino electromagnetic properties at DM experiments. Sec.~\ref{sec:stat} presents the statistical approach, followed by a discussion of our results in Sec.~\ref{sec:res}. We conclude in Sec.~\ref{sec:conc}.

\section{Neutrino electromagnetic properties}
\label{sec:theory}

In this section, we present the relevant \cevns~cross sections in presence of:  neutrino magnetic moments (subsection~\ref{subsec:MM}), neutrino millicharges (subsection~\ref{subsec:EC}), and the possible upscattering of neutrinos into a heavier sterile state (\ref{subsec:DP}). 
All these processes modify the expected \cevns~rate, whose SM cross section is presented in subsection~\ref{subsec:SM}.

\subsection{Standard Model cross section}
\label{subsec:SM}

Within the SM, the \cevns~differential cross section in terms of the nuclear recoil energy, $T_\mathcal{N}$, reads~\cite{Freedman:1973yd,Barranco:2005yy}
\begin{equation}
\label{eq:xsec_CEvNS_SM}
\left. \frac{d\sigma_{\nu_\ell \mathcal{N}}}{dT_\mathcal{N}}\right|^\mathrm{SM}=\frac{G_F^2 m_\mathcal{N}}{\pi}\left({Q_V^\mathrm{SM}}\right)^2 F_{W}^2(\qtransfer^2)\left(1-\frac{m_\mathcal{N} T_\mathcal{N}}{2E_\nu^2} - \frac{T_\mathcal{N}}{E_\nu} \right) \, ,
\end{equation}
where $G_F$ is the Fermi constant, $E_\nu$ the neutrino energy, and $m_\mathcal{N}$ is the mass of the nucleus. The characteristic $N^2$ dependence is encoded in the SM weak charge 
\begin{equation}
\label{eq:CEvNS_SM_Qw}
    Q_V^\text{SM} = g_V^p Z + g_V^n N \, ,
\end{equation}
where $Z~(N)$ stands for the proton (neutron) number of the target nucleus. The tree-level neutron and proton couplings are given by $ g_V^n = -1/2$ and $g_V^p = (1- 4 \sin^2 \theta_W)/2$, where we have fixed the weak mixing angle to $\sin^2 \theta_W=0.23857(5)$~\cite{ParticleDataGroup:2024cfk}. Finite nuclear size effects are taken into account through the nuclear form factor $F_{W}(\qtransfer^2)$, for which we assume the Klein-Nystrand parametrization~\cite{Klein:1999qj} 
\begin{equation}
  \label{eq:KNFF}
   F_{W}(\qtransfer^2)=3\frac{j_1(\qtransfer R_A)}{\qtransfer R_A} \left(\frac{1}{1+\qtransfer^2a_k^2} \right)\ .
\end{equation}
Here, $j_1(x)=\sin(x)/x^2-\cos(x)/x$ is the spherical Bessel function of order one, $a_k = 0.7$~fm and $R_A = 1.23 \, A^{1/3}$ denotes the root mean square radius (in fm), where $A$ is the atomic mass number. 
Note that the typical momentum transfer for \cevns~induced by $^8$B solar neutrinos is $\qtransfer \sim \mathcal{O}(10)$ MeV, so the dependence on the nuclear form factor is expected to be small.

\subsection{Neutrino magnetic moments}

Magnetic moments of neutrinos arise in neutrino mass models and have been constrained by different classes of experiments, see, e.g., Ref.~\cite{Giunti:2024gec}.
Currently, experimental sensitivities are far from reaching the predicted values at the minimal extension of the SM with right-handed Dirac neutrinos~\cite{Fujikawa:1980yx,Pal:1981rm,Shrock:1982sc}. However, in more elaborated models (see, e.g., \cite{Babu:2020ivd}), neutrino magnetic moments might be larger  and could be probed in current and near-future experiments (for a review, see~\cite{Giunti:2014ixa}).

Due to the helicity flipping nature of the interaction induced by the neutrino magnetic moment, its contribution adds incoherently to the SM cross section. Hence, the total cross section is obtained by adding to Eq.~\eqref{eq:xsec_CEvNS_SM} \cite{Vogel:1989iv}

\label{subsec:MM}
\begin{equation}
  \label{eq:xsec_magmom}
  \left. \frac{d\sigma_{\nu_\ell \mathcal{N}}}{dT_\mathcal{N}}\right|^\mathrm{MM}
=
\dfrac{ \pi \alpha^2_\mathrm{EM} }{ m_{e}^2 }
\left( \dfrac{1}{T_\mathcal{N}} - \dfrac{1}{E_\nu} \right)
Z^2 F_{W}^2(\qtransfer^2)
\left| \dfrac{\mu_{\nu_{\ell}}}{\mu_{\text{B}}} \right|^2 \, ,
\end{equation}
where $\alpha_\mathrm{EM}$ is the fine-structure constant and $\mu_B = e/(2m_e)$ is the Bohr magneton. At this point, it should be noted that the effective magnetic moment, $\mu_{\nu_{\ell}}$, depends on the experimental configuration, and special care has to be taken when comparing different bounds on this quantity~\cite{Grimus:2000tq, Grimus:2002vb,Schechter:1981hw,Grimus:2000tq, Miranda:2019wdy,AristizabalSierra:2021fuc}. 
In particular, the effective magnetic moments depend on the entries of the electromagnetic vertex function, which arise from the zero-momentum transfer limit of the magnetic and electric dipole form factors, and on entries of the lepton mixing matrix. They also depend on the typical energy and baseline of the experiment under consideration.

\subsection{Neutrino millicharges}
\label{subsec:EC}

In theories beyond the Standard Model (BSM), such as gauge models that include right-handed neutrinos~\cite{Babu:1989tq}, it is possible for neutrinos to acquire a tiny electric charge (EC).
In principle, neutrinos can have one millicharge corresponding to each neutrino flavor $\ell$, while flavor-changing interactions are also possible. The cross section in the presence of neutrino electric charges is given by~\cite{Chen:2014ypv}
\begin{equation}
\label{eq:xsec_CEvNS_EC_tran}
\left. \frac{d\sigma_{\nu_\ell \mathcal{N}}}{dT_\mathcal{N}}\right|^\mathrm{EC}= \left. \frac{d\sigma_{\nu_\ell \mathcal{N}}}{dT_\mathcal{N}}\right|^\mathrm{SM+EC} + \sum_{\ell'\neq \ell}\frac{G_F^2 m_\mathcal{N}}{\pi}\left({Q_{\ell \ell'}^\mathrm{EC}}\right)^2 Z^2 F_{W}^2(\qtransfer^2)\left(1-\frac{m_\mathcal{N} T_\mathcal{N}}{2E_\nu^2} - \frac{T_\mathcal{N}}{E_\nu} \right) \, ,
\end{equation}
where the first term is obtained from Eq.~\eqref{eq:xsec_CEvNS_SM} and Eq.~\eqref{eq:CEvNS_SM_Qw} through the substitution $g_V^p \to g_V^p - Q_{\ell \ell}^\text{EC}$. This term corresponds to diagonal electric charge interactions that conserve the neutrino helicity and hence interfere coherently with the SM contribution.
Neutrinos could also acquire transition electric charges that induce a flavor change, $\ell \to \ell'$, represented in the second term of Eq.~\eqref{eq:xsec_CEvNS_EC_tran}. In this case, the contribution of the electric charge is added incoherently to the SM term~\cite{Cadeddu:2020lky}.
In both cases, $Q_{\ell \ell^{(')}}^\text{EC}$ is defined as~\cite{Kouzakov:2017hbc,AtzoriCorona:2022qrf} 

\begin{equation}
  \label{eq:QW_millicharge}
 Q_{\ell \ell^{(')}}^\text{EC}  = \frac{2 \sqrt{2} \pi \alpha_\text{EM}}{G_F q^2} \mathcal{Q}_{\nu_{\ell \ell^{(')}} }\, ,
\end{equation}
with $q^2 =-\qtransfer^2= -2 m_\mathcal{N} T_\mathcal{N}$ being the four-momentum transfer and $\mathcal{Q}_{\nu_{\ell \ell^{(')}}}$ the flavor conserving (non-conserving) electric charge.

\subsection{Sterile dipole portal}
\label{subsec:DP}

Several BSM models that seek to explain neutrino masses extend the SM particle content by introducing new heavy neutral leptons~\cite{Abdullahi:2022jlv,Minkowski:1977sc,Yanagida:1979as,GellMann:1980vs,Mohapatra:1979ia,Schechter:1980gr}. Although in principle there is no limit on the number and mass of these sterile states, their phenomenology varies depending on the specific BSM framework considered. Interestingly, these heavy neutral leptons may act as portals to a dark sector or, generically, to new physics at higher energy scales. 
Here we assume that the SM particle content is extended by adding one heavy neutral lepton, $N_4$, with mass $m_4$, which interacts with active neutrinos through the so-called \textit{sterile dipole portal}~\cite{McKeen:2010rx,Magill:2018jla}. This scenario is  characterized by the following Lagrangian\footnote{Note that in the literature the neutrino transition magnetic moment is also expressed as $d_\ell = \frac{\sqrt{\pi \alpha_\mathrm{EM}}}{m_e}\left| \dfrac{\mu_{\nu_{\ell}}}{\mu_{\text{B}}} \right|^2 $ and expressed in units of GeV$^{-1}$, being connected to a new physics scale.}
~\cite{Gninenko:1998nn,Grimus:2000tq}
\begin{equation}
    \mathcal{L}_\mathrm{DP} = \bar{N}_4(i \slashed{\partial}- m_4) N_4 + \frac{\sqrt{\pi \alpha_\mathrm{EM}}}{2 m_e}\left| \dfrac{\mu_{\nu_{\ell}}}{\mu_{\text{B}}} \right|^2 \bar{N}_4 \sigma_{\mu \nu} \nu_\ell F^{\mu \nu} \,, 
\end{equation}
where $F^{\mu \nu}$ is the electromagnetic field tensor, and $\sigma_{\mu \nu} = i (\gamma^\mu \gamma^\nu - \gamma^\nu\gamma^\mu)/2$.
The latter interaction is helicity-flipping and induces an upscattering process of the active neutrino into the sterile state, $\nu_\ell \mathcal{N} \to N_4 \mathcal{N}$, with the mass of the heavy neutral lepton bounded by the kinematic constraint
\begin{equation}
  \label{eq:kinematic_constraint}
  m_4^2\lesssim 2m_\mathcal{N} T_\mathcal{N}\left(\sqrt{\frac{2}{m_\mathcal{N}  T_\mathcal{N}}}E_\nu -1\right)\ .
\end{equation}
The cross section for this process is added incoherently to the SM one and reads~\cite{McKeen:2010rx}
\begin{equation}
  \label{eq:xsec_dipolesterile}
\begin{aligned}
 \left. \frac{d\sigma_{\nu_\ell \mathcal{N}}}{d T_\mathcal{N}}\right|^\mathrm{DP} = &
  \dfrac{ \pi \alpha^2_\mathrm{EM} }{ m_{e}^2 }\, Z^2 F_{W}^2(\qtransfer^2)
\left| \dfrac{\mu_{\nu_{\ell}}}{\mu_{\text{B}}} \right|^2 \\
 & \times \left[\frac{1}{T_\mathcal{N}} - \frac{1}{E_\nu} 
    - \frac{m_4^2}{2E_\nu T_\mathcal{N} m_\mathcal{N}}
    \left(1- \frac{T_\mathcal{N}}{2E_\nu} + \frac{m_\mathcal{N}}{2E_\nu}\right)
    + \frac{m_4^4(T_\mathcal{N}-m_\mathcal{N})}{8E_\nu^2 T_\mathcal{N}^2 m_\mathcal{N}^2}
  \right]  \,  .
  \end{aligned}
\end{equation}
Notice that in the limit $m_4\to 0$, this cross section becomes identical to Eq.~\eqref{eq:xsec_magmom}. The above cross section corresponds to a spin-1/2 nuclear target, although the difference compared to the case of a spin-zero target is negligible~\cite{Miranda:2021kre}. We have also ignored tiny contributions to the cross section arising from the interference between magnetic and weak interactions~\cite{Grimus:1997aa,Miranda:2021kre}.

\section{Statistical analysis}
\label{sec:stat}

We now detail our statistical analysis of the recent PandaX-4T~\cite{PandaX:2024muv} and XENONnT~\cite{XENON:2024ijk} data,  for which we follow the procedure presented in Ref.~\cite{DeRomeri:2024iaw}.
We first estimate the differential number of events as a function of the nuclear recoil energy through the convolution of the neutrino flux with the cross section
\begin{equation}
\label{eq:diffrate}
    \dfrac{dR^\mathrm{X,P}}{dT_\mathcal{N}} = \mathcal{E}^\mathrm{X,P} \, \mathcal{A}^\mathrm{X,P}(T_\mathcal{N})\int dE_\nu \sum_\ell
    \dfrac{d\phi^0}{dE_\nu}~\mathcal{P}(\nu_e\to\nu_\ell)
    \dfrac{\d \sigma_{\nu_{\ell} \mathcal{N}}}{\d T_\mathcal{N}}\,,
\end{equation}
where the exposures are $\mathcal{E}^\mathrm{X} = 3.51$ t$\times$y (X in the superscript is for XENONnT) and $\mathcal{E}^\mathrm{P} = 1.04$ t$\times$y (P for PandaX-4T), while $\mathcal{A}^\mathrm{X,P}(T_\mathcal{N})$ stands for the signal efficiency given in Refs.~\cite{XENON:2024ijk,XENON:2024hup} and~\cite{PandaX:2024muv} for XENONnT and PandaX-4T, respectively.
Moreover, $\dfrac{\d \sigma_{\nu_{\ell} \mathcal{N}}}{\d T_\mathcal{N}}$ indicates any of the \cevns~cross sections given in Sec.~\ref{sec:theory}, while $\dfrac{d\phi^0}{dE_\nu}$ denotes the $^8$B neutrino flux~\cite{bahcall_web,Bahcall:1996qv}, for which we fix the normalization factor $\Phi_\nu^\mathrm{^8B} = 5.46 \times 10^6~\mathrm{cm^{-2}~s^{-1}}$~\cite{Vinyoles:2016djt}.  We account for neutrino oscillations in the Sun through the probability function $\mathcal{P}(\nu_e\to\nu_\ell)$, where we fix the neutrino oscillation parameters to the best fit values given in Ref.~\cite{deSalas:2020pgw}, except for the atmospheric mixing angle, for which we assume maximal mixing, as allowed from current data~\cite{deSalas:2020pgw}. In this case, the oscillation probabilities are given by 

\begin{equation}
    \mathcal{P}(\nu_e\to\nu_\mu) = \mathcal{P}(\nu_e\to\nu_\tau) = 0.5~\left[1-\mathcal{P}(\nu_e\to\nu_e)\right]\,.
\end{equation}
Hence, the flux of $\nu_\mu$ and $\nu_\tau$ at Earth is expected to be the same,  leading to the same bounds for electromagnetic properties related to $\nu_\mu$ or $\nu_\tau$. 

It should be emphasized that the new physics discussed here might affect the propagation of neutrinos inside the Sun. However, these effects are negligible compared to standard matter effects, which are relevant for the neutrino energies of interest here~\cite{Wolfenstein:1977ue,Mikheyev:1985zog}.  
In the case of magnetic moments, it has been shown that the influence of the solar magnetic field on the LMA solar neutrino solution can be ignored without concern~\cite{Barranco:2002te}.
In astrophysical environments like the Sun, electrical neutrality ensures that proton and electron potentials cancel out for electrically charged neutrinos, leaving the matter effect unchanged.
Therefore, we assume standard matter effects in determining the oscillation probability as neutrinos propagate through the Sun.

Note that for the analysis of XENONnT data, we rely on the combination of the S1 (scintillation photons) and S2 (ionization electrons) signals, while for PANDAX-4T we focus on the unpaired S2 (US2)-only signal, as the available information in the experimental paper~\cite{PandaX:2024muv} does not allow for a consistent reproduction of the results of the paired data.
The experimental data are reported in terms of the number of photoelectrons ($n^\mathrm{X}_\text{PE}$) in the case of XENONnT, or the number of electrons ($n^\mathrm{P}_{e^-}$) for PandaX-4T. We thus translate the nuclear recoil energy into the relevant experimental quantity through the following relations

\begin{eqnarray}
\label{eq:change_var}
    n^\mathrm{X}_\text{PE} &= T_\mathcal{N}Q_y^\mathrm{X}(T_\mathcal{N}) g_2\,,\\
    n^\mathrm{P}_{e^-}  &= T_\mathcal{N}Q_y^\mathrm{P}(T_\mathcal{N})\, ,
\end{eqnarray}
where $g_2 = 16.9$ PE/electron. The charge yields $Q_y^\mathrm{X,P}(T_\mathcal{N})$ are taken from Refs.~\cite{XENON:2024xgd,PandaX:2024muv}. We also include some bin-dependent correction factors~\cite{DeRomeri:2024iaw}, which were needed to better reproduce the SM results of the collaborations.
Then, the total predicted number of events in the $i$-th bin is eventually obtained as
\begin{equation}
    N_i^\mathrm{X,P} = R_i^\mathrm{X,P}  + \sum_k B_i^k\,,
\label{eq:pred_Nrate}
\end{equation}
where $R_i^\mathrm{X,P}$ is obtained from Eq.~\eqref{eq:diffrate} after performing the change of variable to the number of (photo-)electrons and integrating over the bin size, $B_i^k$ are the background components from Refs.~\cite{XENON:2024hup,PandaX:2024muv}: accidental coincidence (AC), neutron-related, and electron-recoil (ER) backgrounds for XENONnT, while in the PandaX-4T analysis we include the cathode electrode (CE) 
and micro-discharge (MD) backgrounds.
We compare our predicted number of events with the observed ones using the $\chi^2$ function

\begin{equation}
    \chi^2_\mathrm{X,P}  = \min_{\alpha,\vec{\beta}} \left\{2\left[\sum_k N^\mathrm{X,P} _k - D^\mathrm{X,P} _k + D^{X,P}_k~\ln \left(\frac{D^\mathrm{X,P}_k}{N^\mathrm{X,P}_k}\right)\right] + \left(\frac{\alpha}{\sigma_{\alpha}}\right)^2 + \sum_i \left(\frac{\beta_i}{\sigma_{\beta_i}}\right)^2\right\}\,,
\end{equation}
where $D^\mathrm{X,P}_k$ are the observed numbers of events, $\sum_k D^\mathrm{X}_k = 37$ events in the case of XENONnT for the paired dataset (including both ionization and scintillation signals) and $\sum_k D^\mathrm{P}_k = 332$ for the US2-only signal in PandaX-4T. Moreover, $\alpha$ is a nuisance parameter with $\sigma_\alpha = 12\%$, which accounts for the uncertainty in the prediction for the $^8$B flux~\cite{Vinyoles:2016djt}, while $\beta_i$ and $\sigma_{\beta_i}$ are the nuisance parameters and uncertainties associated with the background components. We assume $\sigma_\mathrm{AC} = 4.8\%$, $\sigma_\mathrm{neutron} =50\%$, $\sigma_\mathrm{ER} = 100\%$, $\sigma_\mathrm{CE} = 31\%$, and $\sigma_\mathrm{MD} = 23\%$, plus a 22\% uncertainty in signal prediction due to data selection and interaction modeling in PandaX-4T. Regarding XENONnT, we also include an uncertainty of 5\% in our signal prediction
related to the fiducial volume. When performing the combined analysis of the XENONnT and PandaX-4T data, we consider the correlated uncertainty on the neutrino flux only once.

\section{Results}
\label{sec:res}

In this section we present the results of our analyses, first for the neutrino magnetic moments, in \ref{subsec:resmagmom}, then for the neutrino millicharges, in \ref{subsec:resEC}, and finally for the sterile dipole portal in \ref{subsec:resDP}.

\subsection{Neutrino magnetic moments}
\label{subsec:resmagmom}

\begin{figure}[!t]
\centering
\includegraphics[width=0.49\textwidth]{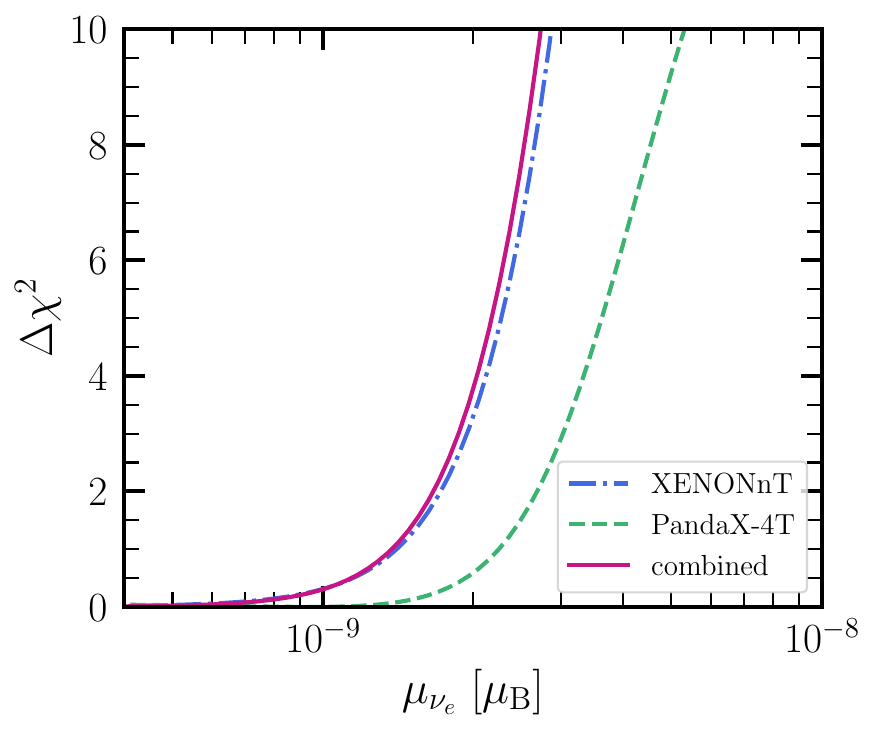}
\includegraphics[width=0.49\textwidth]{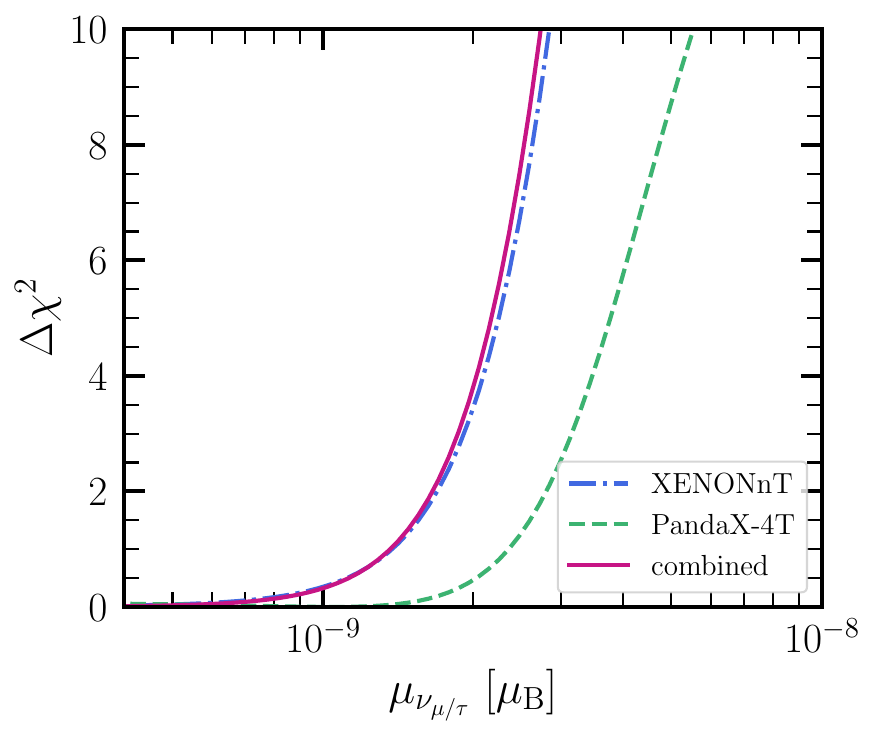}\\
\includegraphics[width=0.49\textwidth]{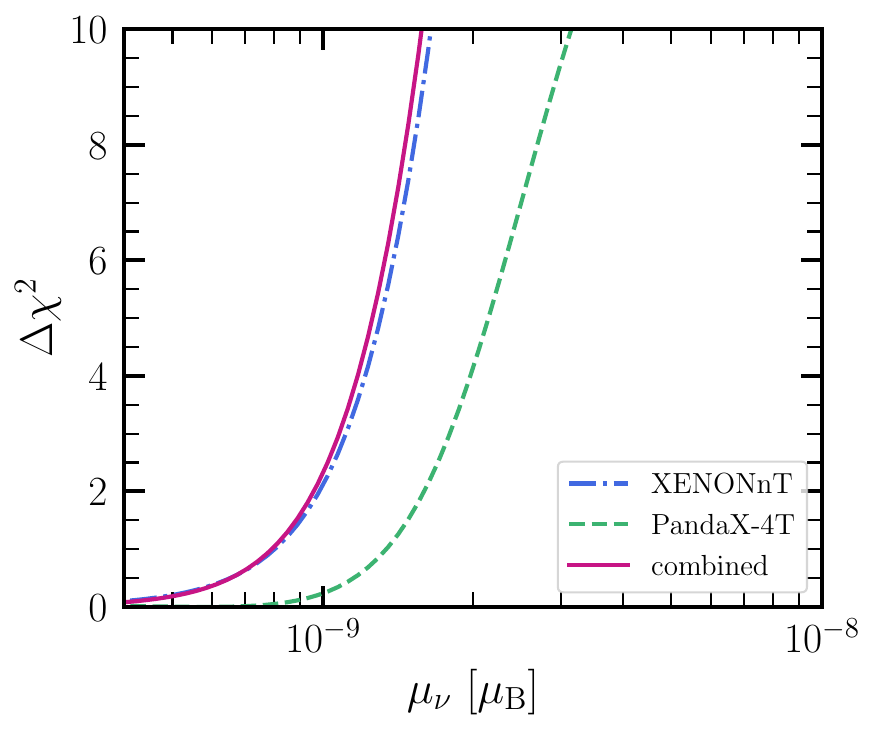}
\caption{Upper panels: $\Delta\chi^2$ profile for the neutrino magnetic moments $\mu_{\nu_{e}}$ and $\mu_{\nu_{\mu, \tau}}$ obtained from the analysis of XENONnT (blue dot-dashed) and PandaX-4T (green dashed) data. The magenta solid line denotes the result for the combined analysis. We assume only one of the magnetic moments to be non-zero at a time.
Lower panel: The corresponding $\Delta\chi^2$ profile assuming $\mu_{\nu} \equiv  \mu_{\nu_{e}} = \mu_{\nu_{\mu}} = \mu_{\nu_{\tau}}$.}
\label{fig:magmom}
\end{figure}

We show in Fig.~\ref{fig:magmom} (upper panels) the $\chi^2$ profiles for the individual effective neutrino magnetic moments, taking only one to be non-zero at a time. 
The green dashed line represents the results from the PandaX-4T analysis, while the blue dot-dashed line corresponds to XENONnT. The solid magenta line illustrates the outcome of the combined analysis, which we find to be entirely dominated by the XENONnT dataset. 
At 90\% confidence level (CL), our obtained bounds read
\begin{align}
\label{eq:magmomres}
     \mu_{\nu_{e}}  <~ & 1.9 \times 10^{-9}~\mu_B \, , \\ 
   \mu_{\nu_{\mu/\tau }} <~ & 1.8 \times 10^{-9}~\mu_B \, .
\end{align}
Recall that the extracted constraints on the muon and tau flavors are identical due to the assumption of maximal atmospheric mixing. Moreover, all three bounds are nearly identical, as expected, since both the survival and appearance oscillation probabilities are approximately 1/3 for neutrino energies relevant to $^8$B neutrinos
For the sake of completeness, we also consider the case of a universal effective magnetic moment, $\mu_\nu \equiv \mu_{\nu_{e}} = \mu_{\nu_{\mu}} = \mu_{\nu_{\tau}}$. The profiles for this analysis are presented in the lower panel of Fig.~\ref{fig:magmom}, while the combined 90\% CL constraint is found to be
\begin{equation}
    \mu_{\nu} < 1.1 \times 10^{-9}~\mu_B\,.
\end{equation}

Notice that the bounds of the present \cevns-based analysis are approximately two orders of magnitude weaker than those derived  using \eves~data from solar neutrinos~\cite{AtzoriCorona:2022jeb,A:2022acy,Giunti:2023yha,XENON:2022ltv,Khan:2022bel}. 
Although the thresholds are similar for both analyses, the SM \eves~cross  section is much smaller than its  \cevns~counterpart. As a result, deviations from the standard scenario become detectable at smaller values of $\mu_\nu$ in the former case. 
As discussed in Sec.~\ref{subsec:MM}, caution is required when comparing bounds from different experiments, since the effective magnetic moment at short baselines differs from that affecting solar neutrinos.
Consequently, these bounds should not be directly compared with those obtained, for instance, from reactor experiments~\cite{TEXONO:2006xds,CONUS:2022qbb,AtzoriCorona:2022qrf}, or COHERENT data~\cite{AtzoriCorona:2022qrf,DeRomeri:2022twg,Coloma:2022avw}. 
For a comprehensive review of all neutrino magnetic moment bounds, including astrophysical limits, see Ref.~\cite{Giunti:2024gec}. 
Let us finally highlight that DM facilities,  sensitive to the solar neutrino flux, can probe the effective magnetic moment for tau neutrinos, which is not accessible instead with experiments using reactor or pion-decaying-at-rest sources.

\subsection{Neutrino millicharges}
\label{subsec:resEC}

\begin{figure}[!t]
\centering
\includegraphics[width=0.49\textwidth]{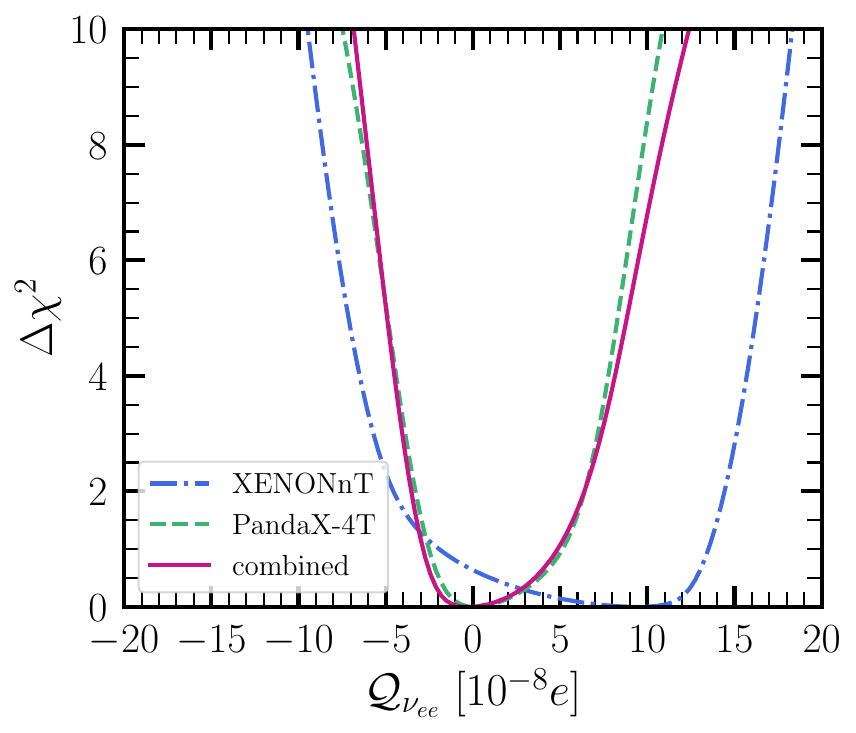}
\includegraphics[width=0.49\textwidth]{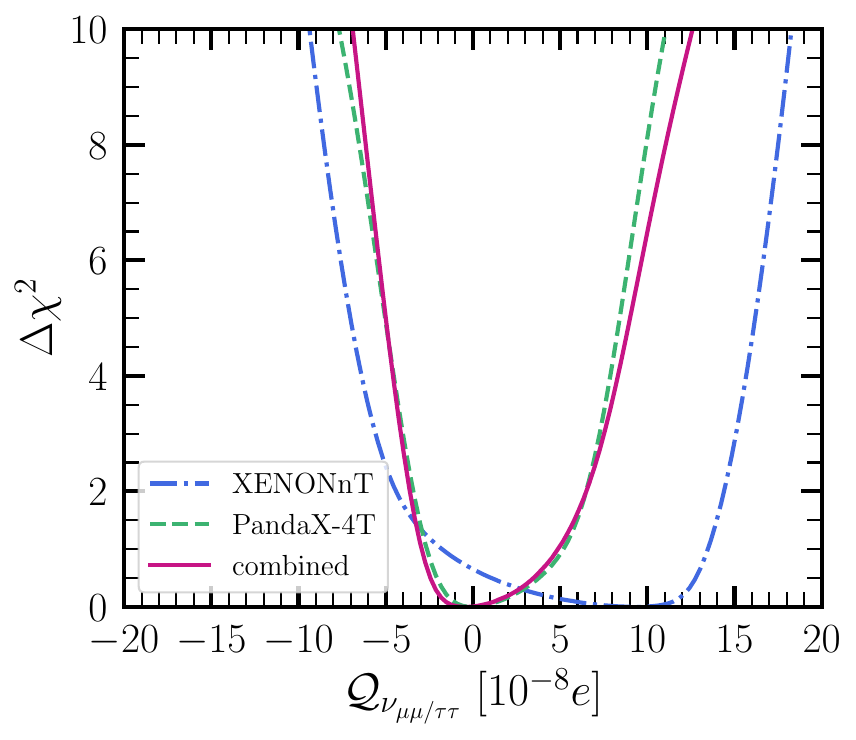}
\includegraphics[width=0.49\textwidth]{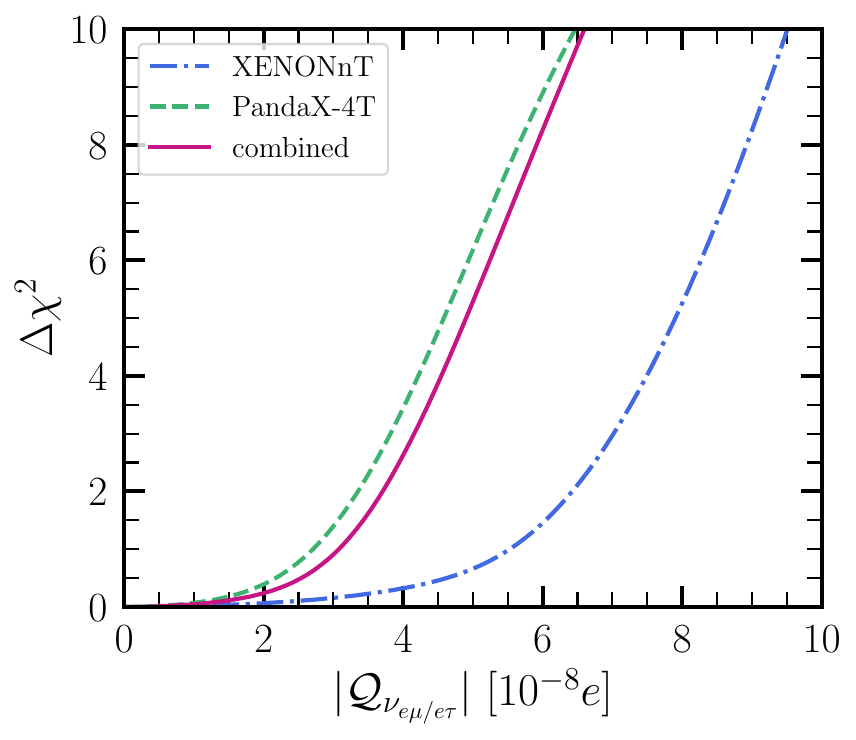}
\includegraphics[width=0.49\textwidth]{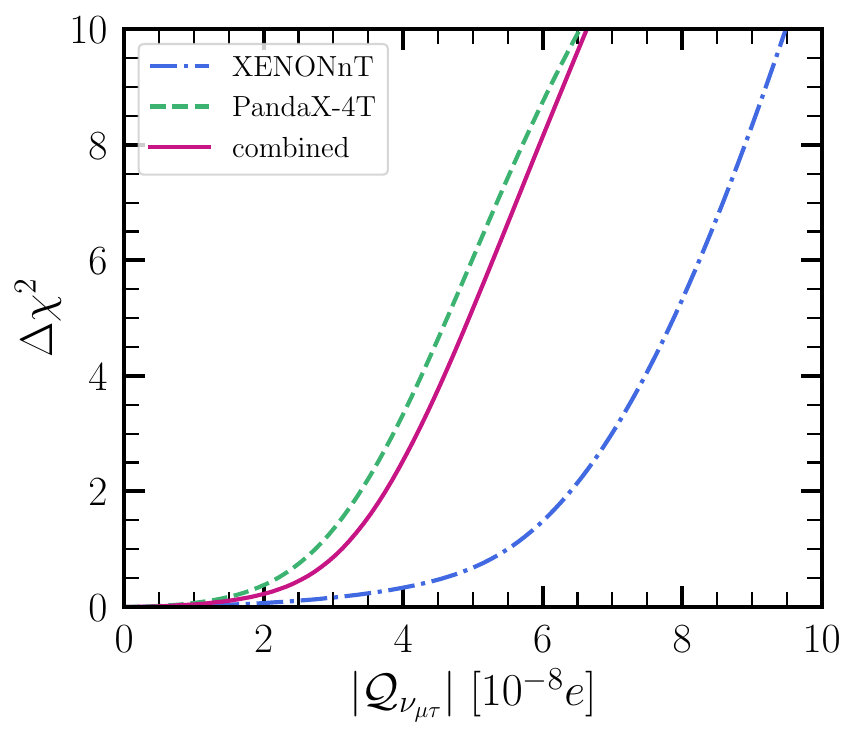}
\caption{$\Delta\chi^2$ profiles
for the flavor-conserving (upper panels) and transition (lower panels) neutrino electric charges.}
\label{fig:millicharge_1D}
\end{figure}

The results obtained for the neutrino electric charges are shown in Fig.~\ref{fig:millicharge_1D}, in terms of $\Delta\chi^2$ profiles for the flavor-conserving (upper panels) and transition (lower panels) millicharges,  with the same color code as in Fig.~\ref{fig:magmom}. 
Unlike in the case of neutrino magnetic moments discussed previously, the neutrino millicharges are correlated, and therefore our present results are obtained by marginalizing over the undisplayed charges.
Therefore, our present results are obtained by marginalizing over the undisplayed charges. 
The combined analysis of XENONnT + PandaX-4T gives, at 90\% CL, 

\begin{align}
\label{eq:ECres}
     \mathcal{Q}_{\nu_{ee}} \in &  \, [-3.9,7.0]\times 10^{-8}~e  \, , \\
    \mathcal{Q}_{\nu_{\mu\mu/\tau\tau}} \in & \, [-4.0,7.2]\times 10^{-8}~e \, , \\
    \left|\mathcal{Q}_{\nu_{e\mu/e\tau}}\right| &  < 4.0\times 10^{-8}~e  \, , \\
     \left|\mathcal{Q}_{\nu_{\mu\tau}}\right| & < 4.1\times 10^{-8}~e  \, .
\end{align}
\begin{figure}[!t]
\centering
\includegraphics[width=\textwidth]{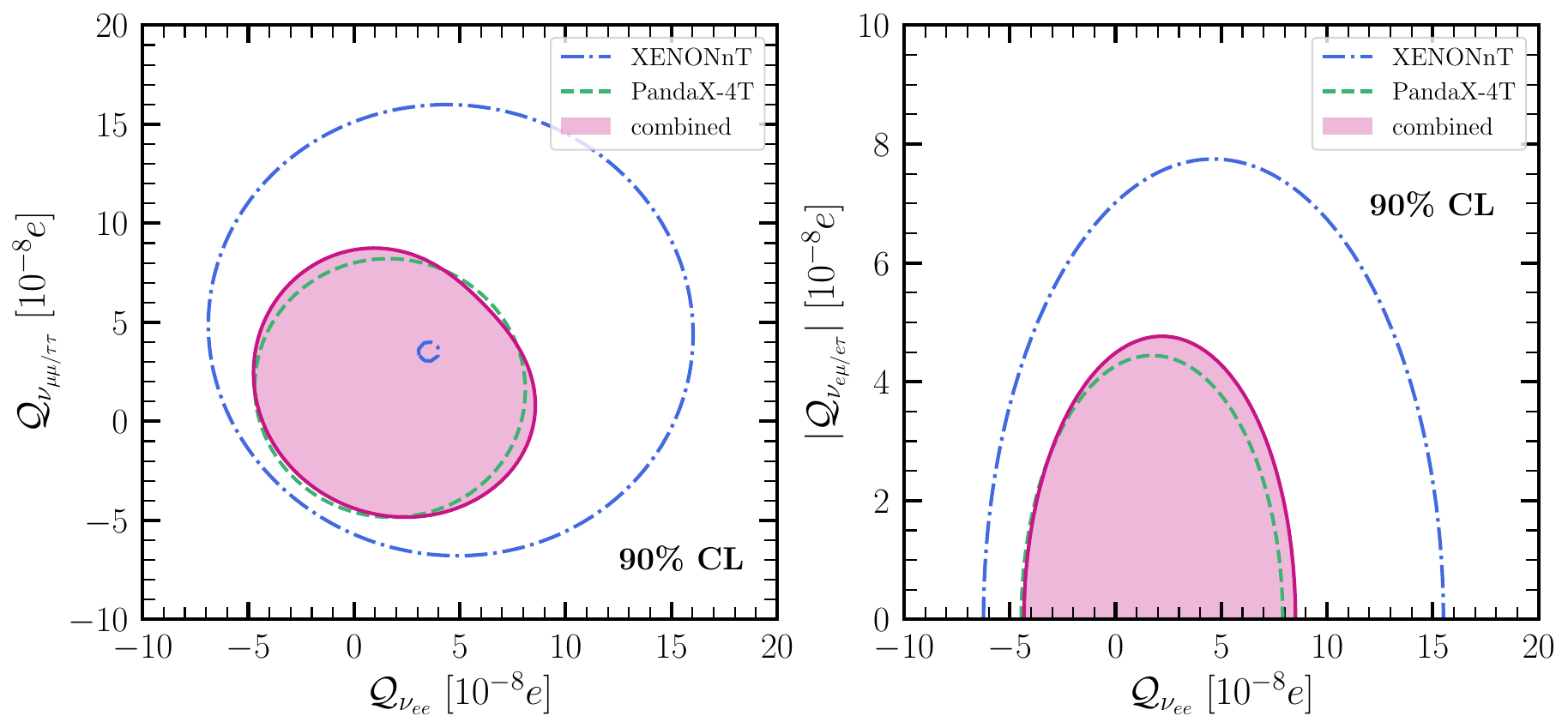}
\caption{Allowed regions at 90 $\%$ C.L. (2 d.o.f.) for the flavor-conserving (left) and a combination of flavor-conserving and flavor-changing (right) neutrino electric charges, in units of the elementary charge $e$.}
\label{fig:millicharge_2D}
\end{figure}
We further show in Fig.~\ref{fig:millicharge_2D} the $90 \%$ CL (2 d.o.f.) allowed regions for the neutrino electric charges, in units of the elementary charge $e$. The left panel corresponds to diagonal charges, while the right panel shows the correlation between $\mathcal{Q}_{\nu_{ee}}$ and $\mathcal{Q}_{\nu_{e\mu /\mu\tau}}$.
As can be seen from the $\Delta\chi^2$ profiles, a non-zero best fit value is found for the flavor-conserving charges in the analysis of XENONnT data. The reason is that, when fixing this value for $\mathcal{Q}_{\ell \ell}$, the predicted number of events in the first (third) $n_\mathrm{PE}^\mathrm{X}$ bin is larger (smaller) than the SM prediction, leading to better agreement with the observed data in the experiment. Note that this behavior was not achieved in the analysis of magnetic moments, where a large $\mu_{\nu_\ell}$ leads to an overall increase of the number of events in all bins.
As a consequence, the XENONnT bounds on electric charges are weaker than the ones obtained from PandaX-4T, in contrast to the case of neutrino magnetic moments. This is also the reason why the combined bound is weaker than the one from PandaX-4T alone. This can be understood from the right panel of Fig.~\ref{fig:millicharge_2D}, where we find that the best-fit value for XENONnT lies outside the allowed region of PandaX-4T, therefore dragging the combined region towards larger values of $\mathcal{Q}_{\nu_{ee}}$.
Overall, and similarly to the magnetic moment case, the present limits are much less stringent than other existing astrophysical and laboratory bounds. In particular, current bounds derived from \eves~and/or \cevns~signals at reactor and spallation source experiments are approximately four orders of magnitude stronger, whereas astrophysical observations lead to even more stringent bounds ~\cite{Giunti:2023yha,Giunti:2024gec,AtzoriCorona:2022qrf,DeRomeri:2022twg}, while the bounds from arguments regarding the neutrality of matter are even stronger~\cite{Giunti:2014ixa} than those.
However, let us finally note that the bounds obtained here could be improved by including \eves~events in the analysis. Indeed, when focusing on (U)S2-only data, both experiments are not capable of distinguishing nuclear from electron recoils. Although subdominant in the SM, the \eves~cross section is sizeably enhanced in the presence of nontrivial electromagnetic properties and especially for millicharges. However, due to the lack of some technical information (efficiency for electron recoils, translation between the number of (photo)electrons and $T_\mathrm{er}$, etc.) it is not possible to include this data sample in analyses carried out outside of the experimental collaboration. 
We leave this exploration for a future work.

\subsection{Sterile dipole portal}
\label{subsec:resDP}

\begin{figure}[!t]
\centering
\includegraphics[width=0.49\textwidth]{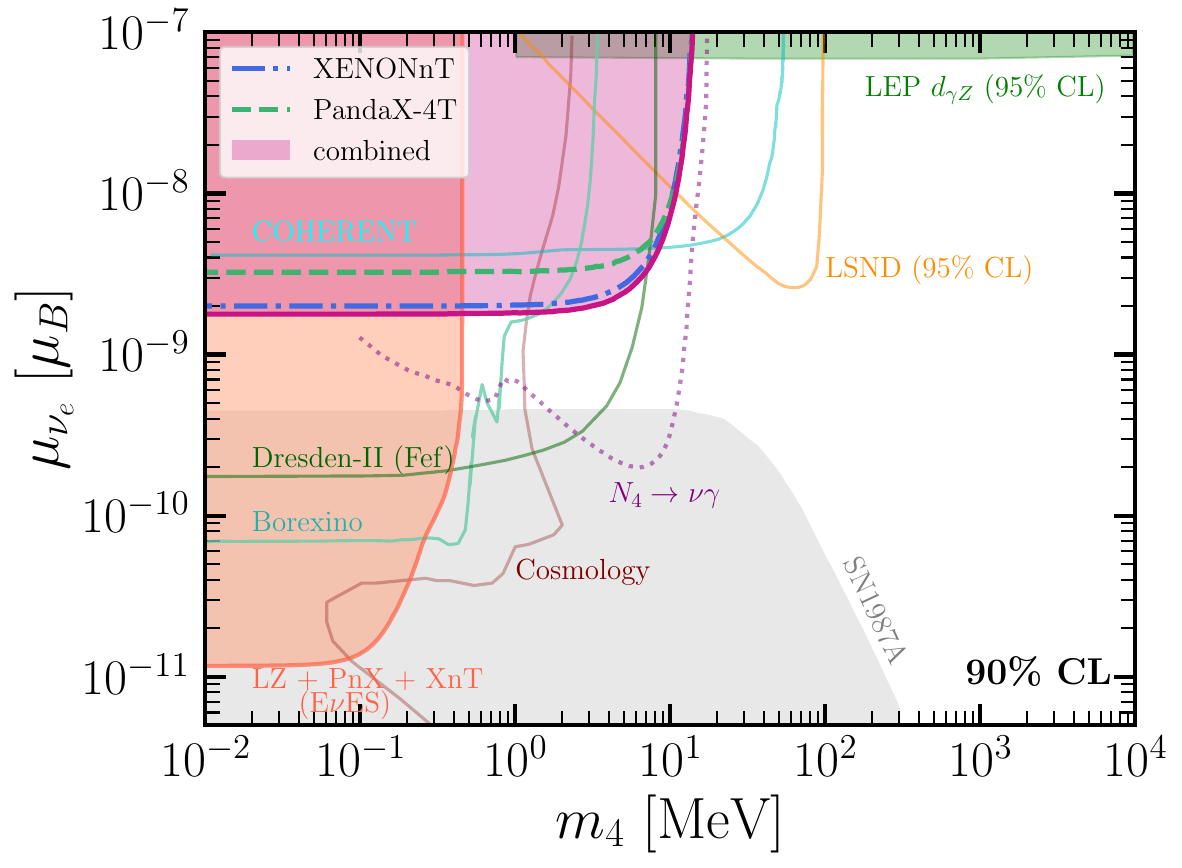}
\includegraphics[width=0.49\textwidth]{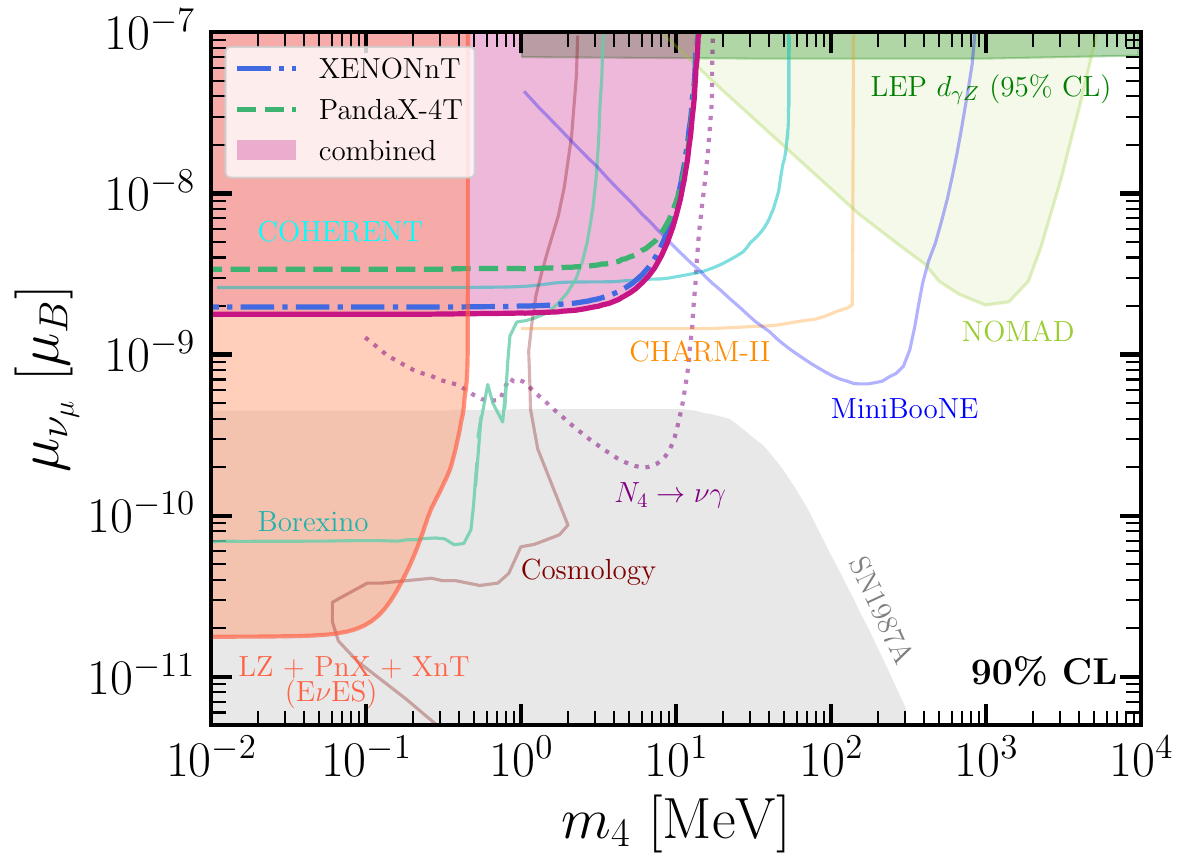}\\
\includegraphics[width=0.49\textwidth]{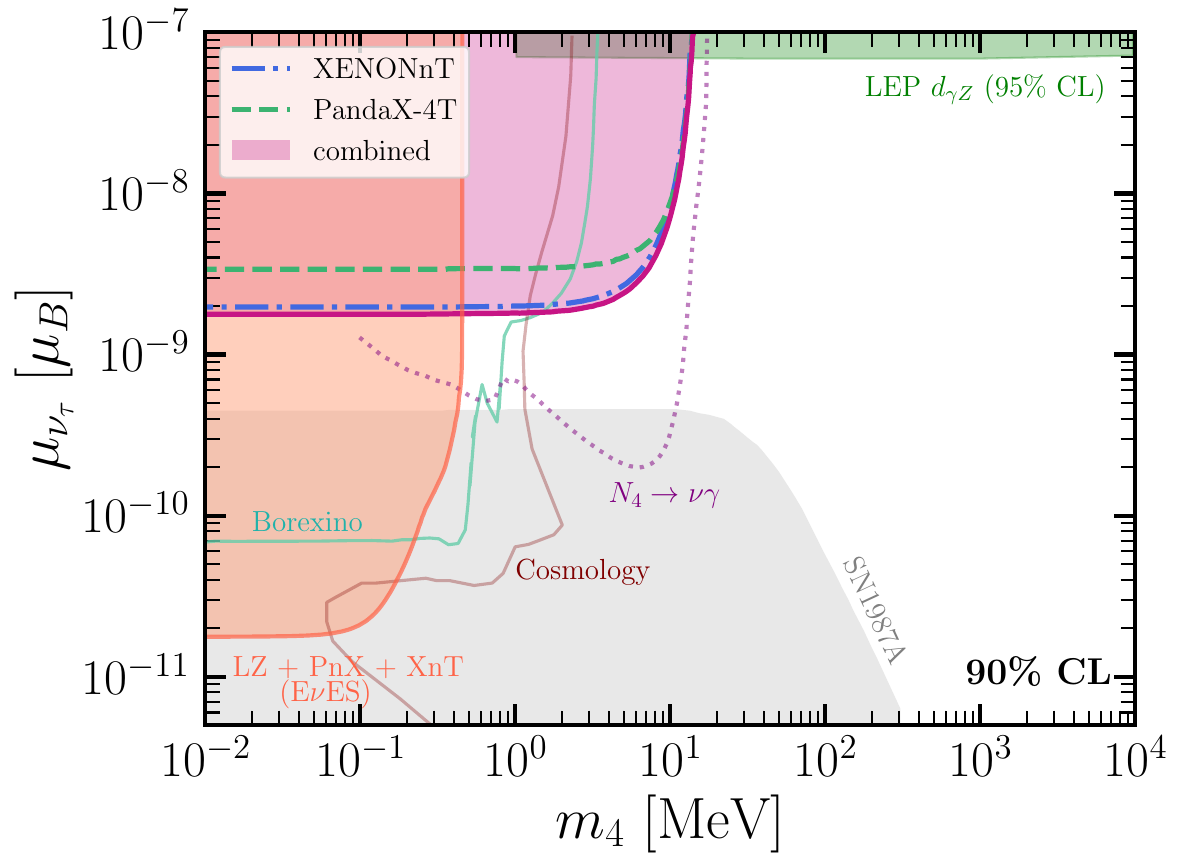}
\caption{$90 \%$ C.L. (2 d.o.f.) exclusion regions for  neutrino magnetic moments inducing  active-sterile transitions for electron neutrinos (upper left panel) for muon neutrinos (upper right panel) and tau neutrinos (lower panel). Results from our analysis of XENONnT~\cite{XENON:2024ijk} and PandaX-4T~\cite{PandaX:2024muv} \cevns~data, and from the combination of them, are shown in blue, green, and magenta, respectively. In coral, we highlight the novel result from a combined analysis of \eves~data at XENONnT~\cite{XENON:2022ltv}, PandaX-4T~\cite{PandaX:2022ood}, and LZ~\cite{LZ:2022lsv}.}
\label{fig:sterile_dipole}
\end{figure}

We finally discuss the results for the sterile dipole portal. We show in Fig.~\ref{fig:sterile_dipole} the 90\% C.L. exclusion contours, in terms of the effective magnetic moment and the mass of the sterile state, $m_4$, produced in the upscattering process for each flavor individually. Given the typical energies of $^8$B solar neutrinos, kinematical constraints allow only to probe sterile states with masses $m_4\lesssim 10$ MeV.  For small masses, $m_4 \lesssim 5$ MeV, the cross section becomes identical to the one of the magnetic moment, see Eq.~\eqref{eq:xsec_magmom}, and hence we recover the bounds from Sec.~\ref{subsec:MM}. We also compute, for the first time, the contours using electron recoil data from XENONnT~\cite{XENON:2022ltv}, LUX-ZEPLIN (LZ)~\cite{LZ:2022lsv}, and PandaX-4T~\cite{PandaX:2022ood}, following the analysis strategy detailed in~\cite{A:2022acy,Giunti:2023yha} which, as can be seen, provides stronger bounds for very small $m_4$.
To complete the picture over the sterile dipole scenario, we also display other existing constraints in Fig.~\ref{fig:sterile_dipole}.  In particular, we compare with bounds from 
COHERENT~\cite{Miranda:2021kre,DeRomeri:2022twg},
Borexino~\cite{Brdar:2020quo,Plestid:2020vqf}, LSND~\cite{Magill:2018jla}, 
LEP~\cite{Magill:2018jla},
MiniBooNE~\cite{MiniBooNE:2007uho},
CHARM-II~\cite{CHARM-II:1989srx,Shoemaker:2018vii},
Dresden-II~\cite{AristizabalSierra:2022axl} (considering the iron-filter (Fef) quenching factor model),
NOMAD~~\cite{NOMAD:1997pcg,Coloma:2017ppo,Magill:2018jla},
$N_4 \to \nu \gamma$ decay~\cite{Plestid:2020vqf},
and from astrophysical and cosmological data such as
SN1987A~\cite{Magill:2018jla,Chauhan:2024nfa}, BBN~\cite{Magill:2018jla,Brdar:2020quo}, and CMB constraints on $\Delta N_\mathrm{eff}$~\cite{Brdar:2020quo} (the latter two denoted as "Cosmology" in the figures).  
To avoid overcrowding the figure, additional bounds and future sensitivities are not shown; instead, we refer the reader to the relevant references~\cite{TEXONO:2009knm,Coloma:2017ppo,Miranda:2019wdy,Schwetz:2020xra,Miranda:2021kre,Bolton:2021pey,AristizabalSierra:2022axl,Gustafson:2022rsz,Barducci:2024nvd,Demirci:2024vzk,Li:2024gbw}.

As shown in the figure, the analysis of recent \eves~data from PandaX-4T, XENONnT, and LZ provides the most stringent laboratory bound in certain regions of the parameter space, and specially at $m_4 \lesssim 0.3$ MeV. For very small masses, the new \eves~constraint is stronger than the Borexino one for all flavors. In a small part of the parameter space it even surpasses the cosmological bound from CMB and BBN. However, we note that this region is already in conflict with the data from SN1987A. On the other hand, the new constraints extracted from the combined analysis of \cevns~data from PandaX-4T and XENONnT, while improving the COHERENT limits below 10 MeV,  cannot (yet) compete with the Borexino and Super Kamiokande limits  from solar neutrino upscattering inside the Earth ($N_4 \to \nu \gamma$), for all flavors. Moreover, for $\nu_\mu$, the bound of CHARM-II is also stronger than the combined PandaX-4T and XENONnT one.
When focusing on \cevns-only experiments, for  the $\nu_e$ active-sterile transition magnetic moment, the Dresden-II limit dominates at small masses, with the present results from PandaX-4T and XENONnT being the strongest in a tiny region of the parameter space, around $m_4 \sim 10$~MeV. For the muon and tau flavors, instead, the present results are the most stringent among all the bounds available from  \cevns~data across most of the parameter space. Specifically, compared to COHERENT, the  bounds obtained from $^8$B solar neutrinos  improve previous limits by a factor 1.5--2 in the mass range $m_4 \lesssim 7$~MeV. At larger masses,  however, they get surpassed by those from spallation-source neutrinos due to kinematic constraints on the upscattered state.

Finally, it is important to note, as with the case of the magnetic moment, that caution is required when comparing limits from different experimental setups, since the effective magnetic moments in different scenarios may not correspond to the same quantities, as explained in Sec.~\ref{subsec:MM}

\section{Conclusions}
\label{sec:conc}
Inspired by indications of \cevns~events induced by $^8$B solar neutrinos observed at two dark matter direct detection experiments, we have investigated how the electromagnetic properties of neutrinos confront these recent results. A non-zero electromagnetic interaction would modify the \cevns~event rate prediction, leading to spectral deviations, particularly enhanced at low recoil energies. Dark matter direct detection facilities have dramatically lowered their recoil energy thresholds, now reaching $T_\mathrm{thr} \sim\mathcal{O}(1)$ keV, making them sensitive to non-negligible rates of $^8$B solar neutrinos inducing \cevns~in their detectors. 
We have analyzed recent data from the XENONnT and PandaX-4T experiments,  considering the electromagnetic properties of neutrinos, specifically focusing on neutrino magnetic moments, millicharges, and the sterile dipole scenario.
Our analysis shows that a combined study  of the XENONnT and PandaX-4T \cevns~data already provides stringent bounds on these quantities. These results are the first results obtained from \cevns~of solar neutrinos on xenon nuclei. Forthcoming data with increased statistics and improved thresholds will allow us to further tighten  these bounds, potentially competing with dedicated terrestrial  facilities. 
Additionally, we note that an analysis focusing on ionization-only data could lead to more stringent bounds through the inclusion of \eves~events.

\section*{Acknowledgments}

C.A.T. is very thankful for the hospitality at IFIC, where part of this work was performed.
We acknowledge useful discussions with Shaobo Wang from the PandaX-4T collaboration.
Work supported by the Spanish grants CNS2023-144124 (MCIN/AEI/10.13039/501100011033 and “Next Generation EU”/PRTR), PID2023-147306NB-I00 and CEX2023-001292-S (MCIU/AEI/ 10.13039/501100011033), as well as by the grants CIDEXG/2022/20, CIAPOS/2022/254 and CIPROM/2021/054 from Generalitat Valenciana.

\bibliographystyle{utphys}
\bibliography{bibliography}  

\end{document}